\documentstyle[aps,pra,epsf,epsfig]{revtex}

\newcommand{\re}{{\rm e}}
\newcommand{\rg}{{\rm g}}
\newcommand{\ra}{{\rm a}}
\newcommand{\rx}{{\rm x}}
\newcommand{\rn}{{\rm n}}
\newcommand{\rs}{{\rm s}}

\title{Driving Atoms Into Decoherence-Free States}
\author{Almut Beige$^{(1)}$, Daniel Braun$^{(2)}$ and Peter L.~Knight$^{(1)}$}
\address{$(1)$ Optics Section, Blackett Laboratory, Imperial College London,
London SW7 2BZ, England. \\
$(2)$ FB7, Universit\"at-GHS Essen, 45117 Essen, Germany.}
\date{\today}

\begin{document}

\maketitle

\begin{abstract}
\begin{center}
\parbox{14cm}{We describe the decoherence-free subspace of $N$ atoms in a
cavity, in which decoherence due to the leakage of photons through the
cavity mirrors is suppressed. We show how the states of the subspace can be
{\em entangled} with the help of weak laser pulses, using the high decay  
rate of the cavity field and strong coupling between the atoms and the 
resonator mode. The atoms remain decoherence-free with a probability which 
can, in principle, be arbitrarily close to unity.}  
\end{center}
\end{abstract}

\vspace*{0.2cm}
\noindent

\section{Introduction}

Following the theoretical formulation of quantum computing \cite{Deutsch86} and 
the first algorithms for problems which can be solved more easily on a 
quantum computer than on a classical computer \cite{Shor:94,Grover97}
the practical implementation of such a device 
has become a challenging task. Initial steps have 
already been taken. Quantum bits (qubits) can be realised for 
instance by storing the information in a superposition of the internal 
states of two-level atoms. To provide the interaction between the atoms 
necessary to perform operations between the qubits the 
coupling via vibrational modes \cite{cp,pcz,molm,schneider} 
or via the single mode inside a cavity \cite{domokos,zoller,rausch} can be used. In other proposals, level shifts due to dipole-dipole 
interaction \cite{brennen,london,briegel} and due to light shifts \cite{jon,steane} have been considered. 

The main limiting factor for quantum computing is decoherence. This normally limits factoring \cite{Shor:94}, for example, to small numbers \cite{Chuang:95,PleKni13} and demonstrates the necessity for error correcting codes \cite{Steane96,Steane962}. But even with the help of quantum error correction, it remains uncertain as to whether decoherence will still destroy the quantum coherence too rapidly for any practical use if the number of qubits required is of the order of
several hundreds or thousands. Indeed, a superposition of two quantum 
mechanical wave functions loses its coherence very
rapidly with the ``distance'' between the components involved \cite{Zurek83}.

However, it has recently become clear that decoherence-free subspaces 
(DFSs) of the total Hilbert space may exist, in which the states are in 
principle exempt from decoherence \cite{Palma:96,Lidar:PRL98,Duan,Zanardi}. 
They arise if the coupling to the environment has a certain symmetry. The 
decoherence-free (DF) states then all acquire the same phase factor, so that
arbitrary superpositions of them remain intact in spite of the interaction
with the environment \cite{Zurek83}. DFSs are promising candidates for quantum 
computing. The dependence of quantum information processing on error
correction schemes is substantially reduced \cite{Lidar}. While the underlying theoretical nature of DFS has received much attention,
far less is known about potential realisations (for examples see Refs.~\cite{Duan33,exps})
and the manipulation of the states {\em inside} the DFS in general 
(see however Refs.~\cite{man,others}). 

\noindent
\begin{center}
\begin{figure}[h]
\epsfig{file=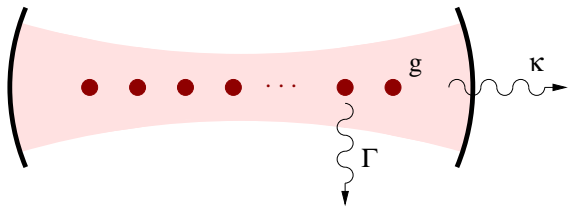,width=7.5cm} \\[0.2cm]
\caption{Schematic view of the system. The two-level atoms are held at fixed
positions in the cavity sufficiently far apart that they can be
addressed individually by laser beams.
}\label{fig.sys}
\end{figure}
\end{center}
\vspace*{0.2cm}

In this paper we give an
example for a DFS which can be implemented using present technology, at
least for small  numbers of qubits and we describe how to prepare and to manipulate the states
{\em inside} a subspace. The system we discuss consists of $N$ macroscopically
separated metastable two-level atoms and is shown in Fig.~\ref{fig.sys}. 
We generate an interaction between
the atoms by placing them at fixed positions in a cavity which acts as a
resonator for an electromagnetic field. The atoms can be stored between the
cavity mirrors 
in a linear trap or in the nodes of a standing light field. The atomic
transition is assumed to be in resonance with a single field mode in the
cavity. The atoms should be strongly coupled to the field mode and the
interaction between each atom with the field is given by the coupling
constant $g_i$. 
As a simplification we assume $g_i \equiv g$ for all $i$, but the 
ideas discussed here can also be carried over to the more general case. 

The main source of decoherence in this system is that a photon can leak 
out through the cavity mirrors with a rate $\kappa$ which is due to 
the coupling of the resonator mode to the free radiation field. Even if 
the cavity mode is empty, the atoms will in general transfer  
excitation into the resonator mode which then can be lost. As we will 
show later this process does not take place if the cavity mode is empty 
and the atoms are prepared in a {\em trapped} state. As a result an example of a DFS is found. The trapped states 
of two two-level atoms in a cavity have been discussed in 
Refs.~\cite{knight,yeoman,alt,meystre}. They belong to a two-dimensional 
Hilbert space which includes the ground state and the maximally entangled 
state $(|1\rangle_1 |0 \rangle_2 -|0 \rangle_1 |1 \rangle_2 )/\sqrt{2}$. 
We will show below that the trapped states of $N$ atoms create a DFS
of dimension  
\begin{eqnarray} \label{15}
{N \choose N/2} ~~&{\rm or}& ~~ {N \choose (N+1)/2}
\end{eqnarray}
for odd and even numbers of atoms, respectively. For large $N$ the dimension
roughly equals $\sqrt{2 / (\pi N)} \cdot 2^N$ and therefore increases with
$N$ almost as fast as the dimension of the whole state space, $2^N$.    

The distance between the atoms should be much larger than an 
optical wavelength. This allows us to address each atom individually
by a single laser pulse. If their Rabi frequencies are much smaller than 
the constants $g$ and $\kappa$, 
laser pulses can be used to prepare and to 
manipulate the states {\em inside} the DFS. 
The reason for this is a  
mechanism which strongly inhibits the transition from trapped to 
non-trapped states in this parameter regime and which can be understood with
the help of the {\em quantum Zeno effect} \cite{misra,itano,behe}. 
We in fact profit from a high decay rate of the resonator
field and the results do not depend on precise values of $g$ and $\kappa$.
Arbitrary
unitary operations can be constructed in a DF qubit formed out of two states
of two atoms. In particular  
we show how a maximally entangled Bell state of the two atoms can be 
generated out of the atomic ground state. 

In the system we discuss here one source of decoherence remains. 
Even if the spontaneous decay rate of the atoms is decreased by the presence 
of the resonator, photons can still be emitted spontaneously into
non-cavity field modes. We therefore propose to use metastable atoms, which
have a very small decay rate $\Gamma$. Spontaneous emission can be neglected
if the duration of the operations performed on the atoms is short compared
to $1/\Gamma$. Therefore the applied laser pulses cannot be arbitrarily
weak, as is necessary for the scheme to work. Care is thus needed to ensure
an overall advantage \cite{PleKni13}. Problems arising from this will be discussed in detail. 

In principle, one could argue that an even larger Hilbert space of 
atomic states than the DFS considered here can be obtained by storing atoms
(or ions) in free space without a surrounding cavity. For this, atomic
decoherence is also due only to spontaneous emission. We should emphasise 
that the major advantage of the system discussed here is that two qubit
entanglement operations can be performed with the help of laser pulses,
while laser pulses cannot entangle atoms in the free space case.  

One method of entangling atoms via their interaction with a resonator mode
is discussed in Ref.~\cite{domokos,rausch} in which the atoms fly through a high
finesse cavity. The time over which the atoms interact with the field is
fixed and determined by the atomic velocity. If the atoms leave the cavity 
their time evolution stops and the prepared state is stable. Using this idea
to perform many operations in a sequence and to scale up the system by using
many atoms 
becomes costly in both time and material. In our approach, the system once
prepared in a state of the DFS, does not change, because the
interaction between the atoms, the cavity mode and the environment of the
system is effectively switched off.  
The atoms can be stored in the cavity over long periods and arbitrarily many 
operations can be performed. 

The paper is organised as follows. In the next Section we give a detailed 
description of the physical system we deal with. In Section III we review
the quantum jump approach \cite{HeWi,HeSo,PleKni} employed to describe the
dissipative dynamics. This approach is equivalent to the Monte-Carlo
wavefunction approach \cite{MC} and to quantum trajectories \cite{QT}. It
also gives a simple criterion for a state to be DF. We construct the
DFS for $N$ atoms in Section IV. How the states in the DFS can be
manipulated is explained in the following two Sections. We summarise our
results in Section VII. 

\section{Description of the physical system}

The system considered here consists of $N$ metastable two-level atoms (or ions)
confined to fixed positions inside an optical cavity. In the following $|0 \rangle_i$ and $|1 \rangle_i$ denote the ground and the excited state of atom $i$, respectively. The Pauli operator $\sigma_i = |0 \rangle_{i\,i}\langle 1|$ is the atomic lowering operator. The atoms with level 
separation $\hbar\omega_0$ are considered to be in resonance with a single mode of the electromagnetic field inside the cavity. The coupling strength for each atom to the cavity mode $g$ is taken as real. The field annihilation operator for the cavity mode is denoted by $b$. 
In addition the atoms are weakly coupled to the free radiation field outside the cavity with a coupling constant $g_{{\bf k} \lambda}^{(i)}$ for the $i$th atom and a field mode with wave vector ${\bf k}$ and polarisation $\lambda$. The annihilation operator for this mode is $a_{{\bf k} \lambda}$. This free radiation field provides an environment for the atoms and is responsible for spontaneous emission. We also take into account non-ideal cavity mirrors by coupling the field inside the resonator to the outside with a strength $\tilde{g}_{{\bf k}
\lambda}$, so that single photons can leak out. The annihilation operator of the free radiation field to which the cavity field couples is given by $\tilde{a}_{{\bf k} \lambda}$. Then, in the Schr\"odinger picture the Hamiltonian of the system and its environment is given by 
\begin{eqnarray} \label{21} 
H &=& \sum_{i=1}^N \hbar \omega_0 \, \sigma_i^\dagger \sigma_i
+ \hbar \omega_0 \, b^\dagger b 
+ \sum_{{\bf k} \lambda} \hbar \omega_k 
\left( a_{{\bf k} \lambda}^\dagger a_{{\bf k} \lambda}
+ \tilde{a}_{{\bf k} \lambda}^\dagger \tilde{a}_{{\bf k} \lambda} \right)
\nonumber \\
& & + {\rm i} \hbar \sum_{i=1}^N g \, b \sigma_i^\dagger 
+ {\rm i} \hbar \sum_{i=1}^N \sum_{{\bf k} \lambda}
g_{{\bf k} \lambda}^{(i)} \, a_{{\bf k} \lambda} \sigma_i^\dagger 
+ {\rm i} \hbar \sum_{{\bf k} \lambda} \tilde{g}_{{\bf k} \lambda} \, 
\tilde{a}_{{\bf k} \lambda} b^\dagger + {\rm h.c.}
\end{eqnarray}
The first four terms give the interaction free Hamiltonian and 
correspond to the free energy of the atoms, the resonant cavity mode and the electromagnetic fields outside the system. Going over to the interaction picture with respect to the interaction free Hamiltonian gives rise to the interaction Hamiltonian
\begin{eqnarray} \label{23} 
H_{\rm I} &=& {\rm i} \hbar \sum_{i=1}^N g \, b \sigma_i^\dagger
+ {\rm i} \hbar \sum_{i=1}^N \sum_{{\bf k} \lambda}
g_{{\bf k} \lambda}^{(i)} \, a_{{\bf k} \lambda} \sigma_i^\dagger 
\, {\rm e}^{{\rm i} (\omega_0-\omega_k) t}
+ {\rm i} \hbar \sum_{{\bf k} \lambda}
\tilde{g}_{{\bf k} \lambda} \, \tilde{a}_{{\bf k} \lambda} b^\dagger 
\, {\rm e}^{{\rm i} (\omega_0-\omega_k) t} + {\rm h.c.}
\end{eqnarray} 
The first term contains the coupling of the atoms to the cavity mode. The second term describes the coupling of the atoms to the free radiation field and is responsible for spontaneous emission with a decay rate $\Gamma$ (see Fig.~\ref{fig.sys}) as will be shown in the next Section. From the last term  the damping of the cavity mode by leaking of photons through the cavity mirrors will arise. The decay rate of a single photon inside the resonator is $\kappa$ and we assume here 
\begin{eqnarray} \label{mag}
g & \sim & \kappa ~,
\end{eqnarray}
i.e.~$g$ and $\kappa$ are of the same order of magnitude. 

To prepare and manipulate the states of the atoms inside the DFS,  
resonant laser pulses are applied, which address each atom
individually. The Rabi frequency of the laser which interacts with atom $i$
will be denoted by $\Omega_i$. The Hamiltonian describing
the effect of the laser in rotating wave approximation and in the
interaction picture chosen above is equal to 
\begin{eqnarray} \label{27} 
H_{\rm laser\,I} &=& {\hbar \over 2} \sum_{i=1}^N \Omega_i \,
\sigma_i + {\rm h.c.}
\end{eqnarray}
We will assume here for all $\Omega_i \neq 0$,
\begin{eqnarray} \label{cond}
\Gamma & \ll & |\Omega_i| ~\ll~ g~.
\end{eqnarray}
Note, that the frequencies $\Omega_i$ are in general complex numbers. Their
phase factors cannot be compensated by changing the basis of the atomic
states, because we have already chosen the coupling constants $g_i$ to be
the same for all atoms. 

To increase the precision of the state preparation, 
detectors could be used which continuously monitor the free radiation field 
outside the system. If a photon is emitted spontaneously or leaks out 
through the cavity mirrors one should stop the experiment and re-initiate the whole process. But even without detectors the experiment can work, in principle, with an arbitrary high success rate. We will show that the probability for the loss of a photon is negligible and only small errors are introduced if it is not recorded.

\section{The conditional time evolution}

One necessary requirement for quantum computing is the ability to manipulate
the qubits in a controlled way. In any quantum algorithm, a system in an
arbitrary pure state has to be transformed into another pure state by
appropriate coherent unitary operations. In general the system considered
here interacts with its environment, stochastically loses a photon and
after a short time has to be described by a density matrix. To avoid this we
consider in the following only the specific time evolution under the
condition that no decay takes place, which can easily be determined from a
quantum jump approach description \cite{HeWi,HeSo} of the system. In this
Section we summarise the main results of this approach. 

With the help of the quantum jump approach a conditional Hamiltonian $H_{\rm cond}$ can be obtained, which describes the time evolution of the system provided no photon is emitted, either by spontaneous emission or by leakage of photons through the cavity mirrors. This Hamiltonian can be evaluated by second order perturbation theory from the expression
\begin{eqnarray}
I \!\! I - {{\rm i} \over \hbar} \, H_{\rm cond} \, \Delta t 
&=& \langle 0_{\rm ph} | \, U_{\rm I}(\Delta t,0) \, |0_{\rm ph} \rangle
\end{eqnarray}
using Eq.~(\ref{23}) and (\ref{27}). Here $|0_{\rm ph} \rangle$ is defined as the vacuum state of the free radiation fields outside the system. In a similar way to that used in  Ref.~\cite{alt}, where the case of two atoms in a cavity was discussed, one finds 
\begin{eqnarray} \label{28}
H_{\rm cond} &=& {\rm i} \hbar \, g \sum_{i=1}^N  b \sigma_i^\dagger 
+ {\rm h.c.} 
- {\rm i} \hbar \, \Gamma \sum_{i=1}^N  \sigma_i^\dagger \sigma_i
- {\rm i} \hbar \kappa \, b^\dagger b + H_{\rm laser \, I}~.
\end{eqnarray}
The corresponding conditional time development operator,
$U_{\rm cond} (t, 0) = \exp ( - {\rm i} H_{\rm cond} t/\hbar )$,
is non-unitary because $H_{{\rm cond}}$ is non-Hermitian. This leads to a decrease of the norm of the vector developing with $U_{\rm cond}$ and is connected to the waiting time distribution for emission of a (next) photon. 
If at $t = 0$ the state of the system is $| \psi_0\rangle$, the state at
time $t$ is given by the normalized state\cite{HeWi,HeSo} 
\begin{eqnarray} \label{210}
|\psi^0 (t) \rangle 
&=& U_{\rm cond}(t, 0) \, |\psi_0 \rangle/\| \cdot \| ~.
\end{eqnarray}
The probability $P_0$ to observe {\em no} photon in $(0,t)$ by a broadband detector (over all space) is 
\begin{equation} \label{29}
P_0 (t,  \psi_0) = \| \, U_{\rm cond} (t,0) \, |\psi_0 \rangle \, \|^2.
\end{equation}

In a real experiment, the emitted photons are actually registered with an efficiency $\eta$ smaller than 1, or even $\eta =0$. Then the system is in case of no photon detection prepared in a statistical mixture of the form
\begin{eqnarray} \label{211}
\left[ P_0 \, |\psi^0 \rangle \langle \psi^0 |
+ (1-\eta)(1-P_0) \, \rho_\bot \right]/ {\rm tr}(\cdot).
\end{eqnarray}
Here $\rho_\bot$ describes the state of the system for the case of photon emissions, which is in general different from the state $|\psi^0 \rangle$ we want to prepare.

\section{Construction of the decoherence-free subspace}

With the help of the quantum jump approach we easily find a necessary and sufficient criterion to establish a decoherence free subspace (DFS). For all states $|\psi \rangle$ of a DFS, the probability for no photon emission for all times $t$ has to remain unity, i.e.
\begin{eqnarray} \label{dfs}
P_0(t,\psi) & \equiv & 1 ~~ \forall ~~ t \ge 0 ~.
\end{eqnarray}
This condition is fulfilled if the system does effectively not interact with
the environment \cite{Lidar:PRL98}. In addition, our criterion demands that the system's own time evolution does not move the state out of the DFS.
In this Section we neglect spontaneous emission $(\Gamma =0)$ and determine all states which fulfill condition (\ref{dfs}). In the following $|{\rm n} \rangle$ denotes a states with $n$ photons in the cavity field mode, $|\varphi \rangle$ corresponds to a state of the atoms only and we define $|{\rm n} \rangle \otimes |\varphi \rangle \equiv |{\rm n} \varphi \rangle$.  

Let us first investigate under what condition the probability density for
the loss of a photon by a system in a state $|\psi \rangle$ is equal to
zero. This  is the case if ${\rm d} P_0(t,\psi)/{\rm d}t |_{t=0}=0$ and leads,
using Eq.~(\ref{210}) and (\ref{29}), to the condition 
\begin{eqnarray} \label{42}
\langle \psi| \left( H_{\rm cond} - H_{\rm cond}^\dagger \right) 
|\psi \rangle &=& 
-2 {\rm i} \kappa \, \langle \psi| \, b^\dagger b \, |\psi \rangle
~=~ 0 ~.
\end{eqnarray}
Therefore each state of the DFS must be of the form
\begin{eqnarray} \label{43}
|\psi \rangle &=& | 0\varphi \rangle ~.
\end{eqnarray}
As expected, only if the cavity mode is empty no photon leaks out through
the resonator mirrors. But condition (\ref{43}) is not yet a sufficient criterion
for the states of a DFS. To assure that $P_0(t,\psi) \equiv 1$ for all times
$t$, the cavity mode must never become populated. All matrix elements of the
conditional Hamiltonian of the form $\langle {\rm n} \varphi' | \, H_{\rm cond} \, | 0\varphi  \rangle $ have to vanish for $n\neq 0$. Using Eq.~(\ref{28}) we
find that this is the case, iff  
\begin{eqnarray} \label{44}
J_- \, |\varphi \rangle &\equiv & 
\sum_{i=1}^N  \sigma_i \, |\varphi \rangle ~=~ 0~.
\end{eqnarray}
Under this condition the system's own time evolution does not drive the state out of the DFS. The states defined by Eq.~(\ref{43}) and (\ref{44}) are also known in the literature as {\em trapped} states  \cite{knight,yeoman,alt,meystre}. An explicit expression for
the trapped states of $N=2, \, 3$ and 4 atoms is given in Ref.~\cite{Duan33}. 

Atomic states which fulfill condition (\ref{44}) are well known in quantum
optics as the Dicke states, of the form $|l,-l\rangle$ in the usual $|j,m
\rangle$ notation \cite{Mandel}. They are eigenstates of the total Pauli
spin operator. The quantum number $l$ can take on the values $1/2$, $3/2,\,.\,.\,.\,$, $N/2$ for $N$ odd and
0, 1,$\,.\,.\,.\,$, $N/2$ for $N$ even. 
The states $|l,-l \rangle$ are highly degenerate, namely 
${N \choose {N/2-l}}- {N\choose {N/2-l-1}}$-fold degenerated 
for $l \le N/2-1$. Together with the single ground state $|N/2,-N/2\rangle$ the 
dimension of the total DFS sums up to the expression given in Eq.~(\ref{15}). 

The Dicke states with a fixed quantum number $l$ are also eigenstates of the
operator $\sum_i \sigma_i^\dagger \sigma_i$ which measures the excitation
$n$ in the system \cite{Mandel}. The relation between $n$ and $l$ is given
by $n=N/2-l$. We describe now how an orthonormal basis for such a subset of states can be found which are orthogonal to all other Dicke states. Using the notation 
\begin{eqnarray} \label{a}
|\ra_{ij} \rangle &\equiv & (|1 \rangle_i |0 \rangle_j 
- |0 \rangle_i |1 \rangle_j )/\sqrt{2}
\end{eqnarray}  
and Eq.~(\ref{44}), it can be proven that each state of the form
\begin{eqnarray}
|\varphi \rangle &=& |0 \rangle_2 \otimes |\ra_{13} \rangle \otimes |\ra_{45}
\rangle \otimes \,. \,. \,. \, \otimes |0 \rangle_N ~,
\end{eqnarray}
in which for instance the first and third atom are in an antisymmetric
state, the second one is in the ground state and so on, is a Dicke
state. Writing down all possible states in which $n$ pairs of atoms are in
the antisymmetric state and all others in the ground state gives a subset of
Dicke states. They all have the same excitation number $n$ and cover
uniformly the whole subspace of Dicke states $|l,-l \rangle$ with
$n=N/2-l$. Now these states can be orthogonalised. An
orthonormal basis for the DFS of $N$ atoms can be obtained by joining
together all atomic subbases for fixed $n$ combined with the
vacuum state of the cavity field. 

Let us define analog to Eq.~(\ref{a}) 
\begin{eqnarray} \label{b}
|\rs_{ij} \rangle \equiv  (|1 \rangle_i |0 \rangle_j 
+ |0 \rangle_i |1 \rangle_j )/\sqrt{2},~ 
|\rg_{ij} \rangle \equiv |0 \rangle_i |0 \rangle_j,~
|\re_{ij} \rangle \equiv |1 \rangle_i |1 \rangle_j
~~& {\rm and} &~~ |{\rm xy} \rangle \equiv |{\rm x}_{12}{\rm y}_{34} \rangle~.
\end{eqnarray}  
Then, for instance, an orthonormal basis of the trapped states of
{\em four} atoms can be obtained by orthogonalising the states 
$|\rg_{12} \, \rg_{34} \rangle$,
$|\rg_{12} \, \ra_{34} \rangle$, 
$|\rg_{13} \, \ra_{24} \rangle$, 
$|\rg_{14} \, \ra_{23} \rangle$, 
$|\rg_{23} \, \ra_{14} \rangle$, 
$|\rg_{24} \, \ra_{13} \rangle$,
$|\rg_{34} \, \ra_{12} \rangle$ and
$|\ra_{12} \, \ra_{34} \rangle$
and one finds
\begin{eqnarray} \label{444}
|\rg \rg \rangle,~ |\rg \ra \rangle,~ 
|\ra \rg \rangle,~ |\ra \ra \rangle,~ 
|x_1 \rangle \equiv (|\rs \rg \rangle - |\rg \rs \rangle)/\sqrt{2} 
~~&{\rm and}&~~ |x_2 \rangle \equiv (|\re \rg \rangle + |\rg \re \rangle 
- |\rs \rs \rangle)/\sqrt{3} ~.
\end{eqnarray}
An orthonormal basis states for the Dicke states of {\em two} atoms is $\{|\rg_{12} \rangle$, $|\ra_{12} \rangle\}$.

In general, to obtain a simple form of the states which form the DF qubits,
one can combine the atoms into pairs. The ground states and the
antisymmetric states of each pair can then form one qubit. Thus for instance the first four states in Eq.~(\ref{444}) could be used to obtain two qubits. In this way we find $N/2$ qubits for an even number of atoms. They belong to a $2^{N/2}$
dimensional subspace of the total DFS. The additional states can serve as auxiliary levels to realise certain logical operations. 

\section{Manipulation of the DF states of two atoms}

We now know how DF qubits can be constructed
resulting from the states of $N$ atoms in a cavity. But to do quantum
computing one also has to be able to perform operations {\em inside} the
DFS. In this Section we discuss using the example of {\em two} atoms how DF states can be manipulated. To do so a weak laser pulse is applied to create Rabi frequencies $\Omega_1$ and $\Omega_2$ which obey condition (\ref{cond}). 
We discuss the effect of the pulse on the system 
with the help of a quantum jump approach description (see
Section III) which also gives the probability for no photon emission, e.g.~the success rate of the proposed experiment.  
It will be shown that the atoms remain DF with a success rate which can, in principle, be arbitrarily close to 1. This is due to a mechanism which decouples trapped states from non-trapped ones, which we will explain in detail. 
A generalisation of the scheme to higher atom numbers is given in the
next Section.

\begin{figure}[htb]
\begin{center}
\epsfxsize10.0cm
\centerline{\epsfbox{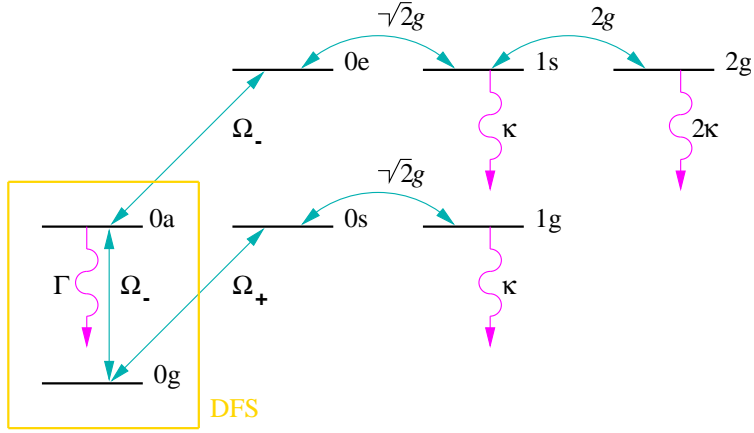}}
\end{center}
\caption{\small Level scheme of the two two-level atoms and the cavity mode
showing the most important possible transitions inside the system. The DFS
contains the states $|0\rg \rangle$ and $|0\ra \rangle$. Two weak lasers
excite the transition inside the DFS and couple it to the states $|0 \re
\rangle$ and $|0\rs\rangle$ with Rabi frequencies $\Omega_-$ and $\Omega_+$,
respectively. Due to the presence of the cavity mode, transitions between
the states $|0\rs \rangle$, $|0\re \rangle$, $|1\rg \rangle$, $|1\rs
\rangle$ and $|2\rg\rangle$ take place with a rate $g$. If the cavity mode
becomes populated a photon can leak out with a rate $\kappa$.} \label{fig1} 
\end{figure}
\vspace*{0.2cm}

In the following we use the same notation as given in Eq.~(\ref{a}) and
(\ref{b}), but suppress  the index 12 for simplicity. As shown above the two
trapped states of two atoms are $|\rg \rangle$ and
$|\ra \rangle$. The states $|\rs \rangle$ and $|\re \rangle$ complete a
basis for the atomic states. 
From Eq.~(\ref{28}) and with the abbreviations 
\begin{eqnarray} \label{36}
\Omega_\pm \equiv (\Omega_1 \pm \Omega_2)/(2\sqrt{2}) 
\end{eqnarray}
the conditional Hamiltonian, which describes the time evolution of the
system under the condition of no photon losses, becomes during the laser
interaction  
\begin{eqnarray} \label{37} 
H_{\rm cond} &=& 
-{\rm i} \hbar g \sum_{\rn=0}^\infty \sqrt{2(n+1)} \, \left( |\rn+1\, g \rangle
\langle \rn\rs|  
+ |\rn+1\, \rs \rangle \langle \rn\re| - {\rm h.c.} \right) \nonumber \\
& & + \hbar \sum_{\rn=0}^\infty \Omega_+ \left( |\rn \rg \rangle \langle \rn\rs| 
+ |\rn\rs \rangle \langle \rn\re| + {\rm h.c.} \right) 
+ \Omega_-  \left( |\rn\rg \rangle \langle \rn\ra| 
- |\rn\ra \rangle \langle \rn\re| + {\rm h.c.} \right) \nonumber \\ 
& & - {\rm i}\hbar  \sum_{\rn=0}^\infty \Gamma
\left( |\rn \ra \rangle \langle \rn\ra| + |\rn\rs \rangle \langle \rn\rs| \right)
+ 2 \Gamma \, |\rn\re \rangle \langle \rn\re|
- {\rm i}\hbar  \sum_{\rn=1}^\infty \sum_{\rx} n \kappa \,
|\rn \rx \rangle \langle \rn\rx|~.
\end{eqnarray}
The first term describes the exchange of excitation between the field mode
and the atoms, while the laser pulses change only the atomic states, as
shown in Fig.~\ref{fig1}. Terms proportional to $\Gamma$ and $\kappa$ are
responsible for a decrease in the norm of the state vector, if higher modes
of the cavity are populated or spontaneous emission of the atoms can take
place. 

Let us assume that the system is in the ground state $|0g\rangle$ at
time $t=0$ when a laser pulse of length $T$ is applied. The {\em
unnormalised} state of the system under the condition of no photon losses
$|\psi^0(t) \rangle$ at time $t$ is denoted in the following by 
\begin{eqnarray}
|\psi^0(t) \rangle &=& \sum_{\rm \rn,\rx} c_{\rm nx}(t) \, |\rn\rx\rangle~.
\end{eqnarray} 
To describe the time evolution of the coefficients $c_{\rm nx}$ we obtain from the time dependent Schr\"odinger equation ${\rm i} \hbar \, {\rm d}/{\rm d}t |\psi^0(t) \rangle = H_{\rm cond} \, |\psi^0(t) \rangle$ a system of differential equations, 
\begin{eqnarray} 
\dot{c}_{\rm ng} &=& -{\rm i} \Omega_- \, c_{\rm na}-{\rm i}
 \Omega_+ \, c_{\rm ns}-\sqrt{2n} \, g\, c_{\rm n-1\,s} -n\kappa \, c_{\rm ng}
\nonumber \\
\dot{c}_{\rm na} &=& -{\rm i} \Omega^{*}_- \, c_{\rm ng}
+{\rm i} \Omega_- \, c_{\rm ne} - (\Gamma+n\kappa) \,  c_{\rm na} 
\nonumber \\
\dot{c}_{\rm ns} &=& -{\rm i} \Omega_+^* \, c_{\rm ng}
- {\rm i} \Omega_+ \, c_{\rm ne} -\sqrt{2n} \, g\,c_{\rm n-1\,e}   
+ \sqrt{2(n+1)} \, g \, c_{\rm n+1 \, g} - (\Gamma+n\kappa) \, c_{\rm ns}
\nonumber \\
\dot{c}_{\rm ne} &=& {\rm i} \Omega_-^* \, c_{\rm na} 
-{\rm i} \Omega_+^* \, c_{\rm ns} + \sqrt{2(n+1)} \, g \, c_{\rm n+1\,s} 
- (2\Gamma+n\kappa) \, c_{\rm ne}~, \label{ampe}
\end{eqnarray}
which will be solved to a good approximation in the following. 

\subsection{Simplified discussion} 

First we discuss the case where the spontaneous emission by the atoms can be neglected and we set $\Gamma=0$. The simplified calculation given in this Subsection describes already the main behaviour of the system due to the laser interaction - the one-qubit rotation. 

As shown in Fig.~\ref{fig1}, only the amplitudes $c_{\rm 0g}$ and 
$c_{\rm 0a}$ change slowly in time, on a time scale proportional to
$1/|\Omega_-|$. Here we are interested in exactly this time evolution. 
All other levels change on a time scale $1/g$ and $1/\kappa$ which is much shorter due to condition (\ref{cond}).
If the system is initially in a DF state the laser pulse excites also the
states $|0 \rs \rangle$ and $|0 \re \rangle$. 
Then the excitation of these levels is transfered with the rate $g$ into
states in which the cavity mode is populated. These states are immediately
emptied by one of the following two mechanisms. One possibility is that a photon leaks out through the cavity mirrors. But,
as long as the population of the cavity field is small, the leakage of a
photon through the cavity mirrors is very unlikely to take place. With a much
higher probability the excitation of the cavity field vanishes during the
conditional time evolution due to the last term in the conditional Hamiltonian in Eq.~(\ref{37}). No
population can accumulate in non-DF states and we can assume $c_{\rm nx}
\equiv 0$ for all states outside the DFS and to zeroth order the differential
equation (\ref{ampe}) simplifies to 
\begin{eqnarray} \label{motion}
\dot{c}_{\rm 0g} &=& -{\rm i} \Omega_- \, c_{\rm 0a} \nonumber \\
\dot{c}_{\rm 0a} &=& -{\rm i} \Omega^{*}_- \, c_{\rm 0g}~.
\end{eqnarray}
This equation describes the time evolution of the DF states to a very good approximation.

If once only the trapped states are populated, the system remains inside the DFS. It behaves like a two-level system with the states $|\rg\rangle$ and $|\ra \rangle$ driven by a laser with Rabi frequency $2\Omega_-$. If the system is initially, when the laser pulse of length $T$ is applied, in the ground state $|0g \rangle$ the atomic state at the end of the pulse, is given by
\begin{eqnarray} \label{324}
|\psi^0(T) \rangle &=& \cos \left(|\Omega_-|T \right) \, |0\rg \rangle 
- {\rm i} \, {\Omega_-^* \over |\Omega_-|} \, \sin \left( |\Omega_-|T \right)
\, |0\ra \rangle. 
\end{eqnarray}
By varying the length $T$ of the laser pulses and control over the phase of $\Omega_-$ any arbitrary rotation between the two states $|0\rg \rangle$ and $|0\ra \rangle$ can be realised. Due to Eq.~(\ref{29}) and (\ref{324}), the probability to find no photon, $P_0(T,\psi_0)$, is unity. Note that the qualitative behaviour is independent of the Rabi frequencies $\Omega_1$ and $\Omega_2$, as long as $\Omega_1 \neq \Omega_2$. To a very good first approximation the atomic states do not move out of the DFS. The quantitative behaviour of the atoms does not depend on the precise values of $g$ and $\kappa$, which simplifies possible realisations of the proposed experiment.

The mechanism which decouples the DFS of the two atoms from the other states
works better, the larger the parameters $g$ and $\kappa$ are compared to
$\Omega_{\pm}$ which is why condition~(\ref{cond}) has been chosen. 
In addition, we assumed $\kappa$ and $g$ to be of the same order of magnitude
(see Eq.~(\ref{mag})) \cite{foot}. Here we use the presence of {\em leaky} cavity mirrors, to ensure that no photon is emitted while the laser pulse is applied! The cavity mode does not become populated during the process which entangles the two atoms with each other and prepares them in the entangled state (\ref{324}). Another example, in which the no-photon time evolution has been used to entangle atoms without a coupling between them via a populated field mode is described in Ref.~\cite{alt}. In Ref.~\cite{bose} it is described how the state of an atom in a cavity can be teleported to an atom inside another distant cavity only by observing emitted photons.

\subsection{A more detailed discussion}

In this subsection we discuss the effect of the laser pulse in more detail and assume again $\Gamma \neq 0$. To solve the differential equations (\ref{ampe}) we make use of an adiabatic elimination suggested by the separation of the frequency scales (\ref{mag}) and (\ref{cond}). Again, Eq.~(\ref{ampe}) shows that the only coefficients that do {\em not} evolve on the fast time scale $g$ or $\kappa$ are $c_{0 \rg}$ and $c_{0 \ra}$. They change with the small rates $\Omega_\pm$ and $\Gamma$. Their time evolution is given by 
\begin{eqnarray} \label{312b}
\dot{c}_{\rm 0g} &=& -{\rm i} \Omega_- \, c_{\rm 0a} 
-{\rm i} \Omega_+ \, c_{\rm 0s} \nonumber \\
\dot{c}_{\rm 0a} &=& -{\rm i} \Omega_-^* \, c_{\rm 0g}
+{\rm i} \Omega_- \, c_{\rm 0e} - \Gamma\, c_{\rm 0a} ~.
\end{eqnarray}
The amplitudes of all other states, which evolve on the fast time scale $g$ or $\kappa$, follow the slowly varying coefficients $c_{0\rg}$ and $c_{0\ra}$. Therefore we can neglect their derivatives compared to the fast rates $g$ and $\kappa$. Setting the derivatives of $c_{0\rs}$, $c_{0\re}$, $c_{1\rg}$, $c_{1\rs}$ and $c_{2\rg}$ in Eq.~(\ref{ampe}) equal to zero we obtain the equations  
\begin{eqnarray} \label{312}
0 &=& -{\rm i} \Omega_+^* \, c_{\rm 0g}
- {\rm i} \Omega_+ \, c_{\rm 0e}    
+ \sqrt{2} \, g \, c_{\rm 1 g} - \Gamma \, c_{\rm 0s}~, \nonumber \\
0 &=& {\rm i} \Omega_-^* \, c_{\rm 0a} 
-{\rm i} \Omega_+^* \, c_{\rm 0s} + \sqrt{2} \, g \, c_{\rm 1s} 
- 2\Gamma \, c_{\rm 0e}~,  \nonumber \\
0 &=& -{\rm i} \Omega_- \, c_{\rm 1a}-{\rm i}
 \Omega_+ \, c_{\rm 1s}-\sqrt{2} \, g\, c_{\rm 0s} -\kappa \, c_{\rm 1g}
\nonumber \\
0 &=& -{\rm i} \Omega_+^* \, c_{\rm 1g}
- {\rm i} \Omega_+ \, c_{\rm 1e} -\sqrt{2} \, g\,c_{\rm 0e}   
+ 2g \, c_{\rm 2 g} - (\Gamma+\kappa) \, c_{\rm 1s}~, \nonumber \\
0 &=& -{\rm i} \Omega_- \, c_{\rm 2a}-{\rm i}
 \Omega_+ \, c_{\rm 2s}-2g\, c_{\rm 1 s} -2\kappa \, c_{\rm 2g}~.
\end{eqnarray}
From Fig.~\ref{fig1} and Eq.~(\ref{ampe}) we can see that all
other coefficients corresponding to non-DF states are smaller by at least one factor of $|\Omega_{\pm}|/g$, because they can only be excited via driving
with the weak laser pulse if the states $|1\rs \rangle$ and $|2\rg \rangle$
are populated. The amplitudes of these higher states can therefore be
neglected in Eq.~(\ref{312}) and we obtain a closed set of equations which can be solved easily for  the coefficients of the DF states. We find 
\begin{eqnarray} \label{315}
\left( \begin{array}{c} 
\dot{c}_{\rm og} \\ \dot{c}_{\rm oa} \end{array} \right)
&=& - \left( \begin{array}{cc}
k_1 & {\rm i} \Omega_- \\ {\rm i} \Omega_-^* & k_2 \end{array} \right)
\left( \begin{array}{c} c_{\rm og} \\ c_{\rm oa} \end{array} \right)
~\equiv~ -M \left( \begin{array}{c} c_{\rm og} \\ c_{\rm oa} 
\end{array} \right)
\end{eqnarray}
with
\begin{eqnarray} \label{316}
k_1 \equiv {|\Omega_+|^2 \kappa \over 2 g^2} ~~&{\rm and}&~~
k_2 \equiv {|\Omega_-|^2 (2g^2+\kappa^2) \over 2 g^2 \kappa} +\Gamma ~.
\end{eqnarray}
The eigenvalues of $M$ are
\begin{eqnarray} \label{317}
\lambda_{1/2} &=& {k_1+k_2 \over 2} \pm {\rm i} |\Omega_-| \,
\sqrt{1-\left( {k_1-k_2 \over 2|\Omega_-|} \right)^2}
~\equiv~ {k_1+k_2 \over 2} \pm {\rm i} S~.
\end{eqnarray}
Making use of the formula 
\begin{eqnarray} \label{318}
{\rm e}^{-Mt} 
&=& {M-\lambda_2 \over \lambda_1-\lambda_2} \,  {\rm e}^{-\lambda_1t}
+ {M-\lambda_1 \over \lambda_2-\lambda_1} \,  {\rm e}^{-\lambda_2t}~,
\end{eqnarray}
which can be checked by applying it to the eigenvectors of $M$ \cite{gant} we find 
\begin{eqnarray} \label{319}
\left( \begin{array}{c} c_{\rm og}(t) \\ c_{\rm oa}(t) 
\end{array} \right)
~=~ {\rm e}^{-Mt} \, \left( \begin{array}{c} 1 \\ 0 \end{array} \right)
&=& {\rm e}^{ - (k_1+k_2)t/2 } \, \left[ 
\left( \begin{array}{c} 1 \\ 0 \end{array} \right) \cos St -{1 \over 2S}
\left( \begin{array}{c} k_1-k_2  \\ 2{\rm i}\,\Omega_- \end{array} \right)
\sin St \right]~,
\end{eqnarray}
which are the coefficients of the DF states at time $T$ under the condition of no photon emission.

After the laser pulse is turned off at time $T$ the excitation of all non-DF states vanishes during a short transition time of the order $1/g$ and $1/\kappa$ due to the conditional time evolution. Therefore the state of the atoms shortly after $T$ and under the condition that no photon was emitted can be
obtained by normalising the state $c_{\rm 0g}(T) \, |0\rg \rangle + c_{\rm
0a}(T) \,|0\ra \rangle$. It equals 
\begin{eqnarray} \label{322}
|\psi^0 (T) \rangle
&=& \left[ \left( \cos ST - {k_1-k_2 \over 2S} \sin ST \right) |0\rg \rangle
- {\rm i} \frac{\Omega_-^*}{S}\sin ST \, |0\ra \rangle \right] / \| \cdot \| ~.
\end{eqnarray}
The probability of a successful operation is given by the probability for no photon emission in $(0,T)$. According to Eq.~(\ref{29}) it is given by $|c_{0\rg}(T)|^2 + |c_{0\ra}(T)|^2$ and leads to
\begin{eqnarray} \label{323}
P_0(T,g) &=& {\re}^{-(k_1+k_2)T} \, \left[ 
1-{k_1-k_2 \over S} \sin ST \cos ST
+ {(k_1-k_2)^2 \over 2S^2} \sin^2 ST \right]~.
\end{eqnarray}

The state $|\psi^0(T) \rangle$ belongs to the DFS. Using Eq.~(\ref{28}), (\ref{43}) and (\ref{44}) one can show $H_{\rm cond} |\psi^0(T)\rangle = 0$ and 
$|\psi^0(T) \rangle$ is now - without the laser interaction - stable in time.
If one neglects again all terms proportional to $\Gamma$ and $|\Omega_\pm|/g$ 
Eq.~(\ref{324}) agrees with the result given in Eq.~(\ref{324}). The laser pulse performs a rotation on the DF qubit. As can be seen from Eq.~(\ref{323}), the sum $k_1+k_2$ can be interpreted as the decay rate of the system. As long as this rate is much smaller than $1/T$ the probability for a successful preparation is close to 1.

\subsection{Preparation of a maximally entangled state of the atoms}

\begin{center}
\begin{figure}[h]
\epsfig{file=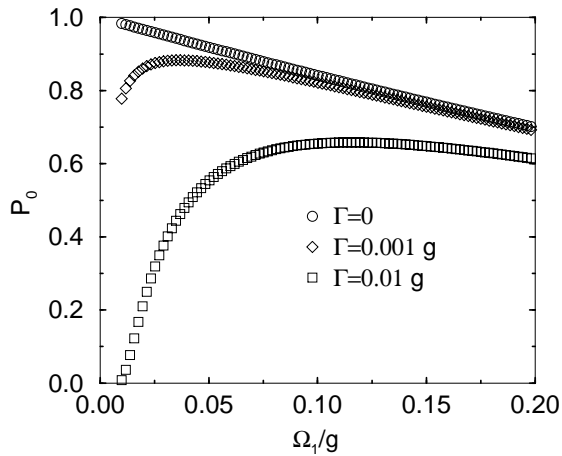,height=7cm} \\[-0.3cm]
\caption{
Probability for successful preparation of the maximally entangled DF state
$|0\ra\rangle$ as a function of the Rabi frequency $\Omega_1$ for
$\Omega_2=-\Omega_1$, $\kappa=g$ and different values of $\Gamma$.
}\label{p0}
\end{figure}
\end{center}

Finally, we discuss as an example the preparation of the maximally entangled atomic state $|\ra\rangle$ while the cavity is empty. Due to Eq.~(\ref{322}) this can be done by choosing the length of the laser pulse equal to 
\begin{eqnarray}
T &=& {1 \over S} \, {\rm arccot} \, {k_1-k_2 \over 2S}  
\approx {\pi \over 2|\Omega_-|}~. 
\end{eqnarray}
Fig.~\ref{p0} shows the success rate $P_0$ for this scheme and results from a numerical solution of Eq.~(\ref{ampe}). The result agrees in the chosen parameter regime very well with $P_0(T,0{\rm g})$ given in Eq.~(\ref{323}).
For zero spontaneous emission, success rates arbitrarily close to unity can be achieved by reducing the Rabi frequency $\Omega_1$. However, for $\Gamma \neq 0$ this is not possible. If the laser pulse becomes very long the probability of occurance of a spontaneously emitted photon increases and is no longer negligible. For finite values of $\Gamma$ there is an optimal value of $\Omega_1$ for which the success rate of the preparation scheme has a maximum.

If all outcoming photons are registered and the experiment is repeated in case of an {\em emission} the fidelity of the prepared state can, for a very wide parameter regime, be very close to 1. For the parameters given in Fig.~\ref{p0} it is always higher than 99$\,\%$. If the photons are registered only with an efficiency $\eta$ smaller than 1, this fidelity has to be multiplied with $P_0/(1-\eta(1-P_0))$ as can be seen from Eq.~(\ref{211}) to then give 
the fidelity of the prepared state in the case of no photon {\em detection}. 

\section{Manipulation of the DFS in general}

In the last Section we have shown that a weak enough laser pulse does not move the state of the system of two atoms out of the DFS. In this Section we want to point out a physical principal behind this fact which allows a straightforward generalisation of the preparation scheme to higher numbers of atoms in the cavity and other kinds of interaction. To do so we shortly review the quantum Zeno effect \cite{misra}. We also derive an {\em effective} Hamiltonian to describe the effect of a weak interaction in general.

The quantum Zeno effect \cite{misra} is a theoretical prediction for the behaviour of a system under rapidly repeated {\em ideal} measurements. It is a consequence of the projection postulate of von Neumann and L\"uders \cite{neu} which describes the effect of a single measurement and predicts that the probability to measure whether the state of a system belongs to a certain subspace of states is given by its overlap with the subspace. If the outcome of the measurement is ``yes'' the state of the system changes during the measurement process. It becomes projected onto the subspace. The quantum Zeno effect predicts that if the time between subsequent measurements equals zero the outcome of each following measurement is the same, even if an additional interaction is applied which is intended to move the system into a complementary subspace. The system can only change {\em inside} the subspace.

We now reconsider the system of $N$ atoms inside the cavity and assume first that no laser pulse is applied to the atoms. Let us define $\Delta T$ as a time, in which a photon is emitted with probability very close to unity, if the system is prepared in a non-DF state. Then the observation of the free radiation field outside the system over a time interval of the length $\Delta T$ can be interpreted as a measurement of whether the system is DF or not. If a photon is emitted, the system has not been in a DF state. Otherwise, its state belongs to the DFS. In the presence of a laser pulse the state of the system can be driven out of the DFS during $\Delta T$, but as long as 
\begin{equation} \label{dt2}
|\Omega_i| \ll 1/\Delta T
\end{equation}
this effect can be negelected and the observation of the free
radiation field over a time interval $\Delta T$ can still be interpreted as
a measurement of whether the atoms are DF or not to a very good
approximation. This is the case in the scheme we discuss here. As it has been shown in the previous Section, $\Delta T$ has
to be at least of the order $1/g$ and $1/\kappa$ and condition (\ref{dt2})
leads to condition (\ref{cond}) given in the Introduction.

In the scheme we propose the free radiation field outside the cavity is observed continuously, i.e., the time between two subsequent measurements is zero. Therefore the quantum Zeno effect can be used to predict the effect of the laser pulse on the time evolution of the system. It suggests, that the system always remains DF if it is once prepared in a state of the DFS.

A generalisation of the proposed scheme to other forms of state manipulation is
straightforward. As long as the interaction is weak enough the state of the system does not move out of
the DFS. The interpretation of the behaviour of the system with the help of the quantum Zeno effect can also be used to derive an {\em effective} Hamiltonian $H_{\rm eff}$ which describes the effect of a weak laser pulse on the system. We know that the state of the system can change only {\em inside} the DFS due to  rapidly repeated measurements whether the system is still DF. Therefore the time development operator for a short time interval $\Delta T$ is to a good approximation given by
\begin{eqnarray} 
U_{\rm eff} (\Delta T,0) 
&=& I\!\!P_{\rm DFS} \, U_{\rm cond}(\Delta T,0) \, I\!\!P_{\rm DFS}~,
\end{eqnarray}
where $I\!\!P_{\rm DFS}$ is the projector onto the DFS. This leads to the effective Hamiltonian
\begin{eqnarray} 
H_{\rm eff} &=& I\!\!P_{\rm DFS} \, H_{\rm cond} \, I\!\!P_{\rm DFS}~.
\end{eqnarray}
If we assume that spontaneous emission by the atoms is negligible $(\Gamma=0)$ the definition of the DF state by Eq.~(\ref{43}) and (\ref{44}) allows to simplify this equation. From Eq.~(\ref{28}) we find
\begin{eqnarray} \label{bla} 
H_{\rm eff} &=& I\!\!P_{\rm DFS} \, H_{\rm laser\,I} \, I\!\!P_{\rm DFS}~,
\end{eqnarray}
where $H_{\rm laser\,I}$ describes the laser interaction and is given in  Eq.~(\ref{27}). The effect of the laser on the system considered here is very different from its effect on atoms in free space. It confines the system inside the DFS and can be used to generate entanglement between the atoms in the cavity. The effective Hamiltonian for a single laser pulse depends on $N$ different Rabi frequencies which can be chosen arbitrarily. This allows to perform a wide range of operations like the CNOT quantum gate between the qubits of a DFS. A concrete proposal for quantum computation using dissipation which is based on the idea discussed here in detail can be found in Ref.~\cite{CNOT}.

In the case of {\em two} atoms which has been discussed in the previous Section the effective Hamiltonian (\ref{bla}) equals 
\begin{eqnarray}
H_{\rm eff} &=& {\hbar \over 2 \sqrt{2}} \, (\Omega_1-\Omega_2) \, |0\rg \rangle \langle 0\ra| + {\rm h.c.} 
\end{eqnarray}
and leads directly to Eq.~(\ref{motion}) in the previous Section. The DFS of {\em four} atoms is six dimensional. Using the notation given in Eq.~(\ref{444})  we find 
\begin{eqnarray}
H_{\rm eff} &=& {\hbar \over 2 \sqrt{2}} \, \Bigg[ \, 
(\Omega_1+\Omega_2-\Omega_3-\Omega_4) 
\left( {1 \over \sqrt{2}} \,|0\rg\rg \rangle \langle 0{\rm x}_1| 
+ \sqrt{2 \over 3} \,|0{\rm x}_1 \rangle \langle 0{\rm x}_2| \, \right)
+ {\rm h.c.} \nonumber \\
& & + (\Omega_1-\Omega_2) 
\left( |0\rg\rg \rangle \langle 0\ra\rg|
+ |0\rg\ra \rangle \langle 0\ra\ra|
-{1 \over \sqrt{3}} \,|0\ra\rg \rangle \langle 0{\rm x}_2| \, \right)
+ {\rm h.c.} \nonumber \\
& & + (\Omega_3-\Omega_4) 
\left( |0\rg\rg \rangle \langle 0\rg\ra|
+ |0\ra\rg \rangle \langle 0\ra\ra|
-{1 \over \sqrt{3}} \,|0\rg\ra \rangle \langle 0{\rm x}_2| \, \right)
+ {\rm h.c.} \, \Bigg]~.
\end{eqnarray}

\section{Conclusions}

In summary, we have given an example of a DFS suitable
for quantum computing and have identified a mechanism for the manipulation
of states {\em within} the DFS which can be understood in
terms of the quantum Zeno effect and allows for generalisation to other
forms of manipulation. This concept was demonstrated in detail for the example of two two-level atoms, which lead to an efficient method of entangling them and was genereralized to $N$ two-level atoms.  

{\em Acknowledgement:}
We thank F.~Haake, W.~Lange, R.~Laflamme, D. A.~Lidar, M. B.~Plenio, B. Tregenna and T.~Wellens for fruitful discussions. Part of this work has been done at
the ESF-Newton Institute Conference in Cambridge and was partially supported
by the ESF. This work was also supported by the A.~v.~Humboldt Foundation, by the European Union, by the UK Engineering and Physical Sciences Research Council and by the Sonderforschungs\-be\-reich 237 ``Unordnung und gro{\ss}e Fluktuationen".


\begin{references}
\bibitem{Deutsch86} D.~Deutsch, Proc.~R.~Soc.~A {\bf 400}, 97 (1985); {\em
ibid.} {\bf 425}, 73 (1989).
%
\bibitem{Shor:94} 
P. W. Shor, {\em Algorithms for Quantum Computation: Discrete Log and Factoring}, eds by S. Goldwasser, Proceedings of the 35th Annual Symposium on the Foundations of Computer Science, IEEE Computer Society,  Los Alamitos, CA (1994), p. 124.
%
\bibitem{Grover97} 
L. K. Grover, Phys. Rev. Lett. {\bf 79}, 325 (1997).
%
\bibitem{cp} 
J. I. Cirac and P. Zoller, Phys. Rev. Lett. {\bf 74}, 4091 (1995).
%
\bibitem{pcz}
J. F. Poyatos, J. I. Cirac and P. Zoller, Phys. Rev. Lett. {\bf 81}, 1322 (1998). 
%
\bibitem{molm} 
A. S{\o}rensen and K. M{\o}lmer, Phys. Rev. Lett. {\bf 82}, 1971 (1999);
A. S{\o}rensen and K. M{\o}lmer, Phys. Rev. A (in press), quant-ph/0002024. 
%
\bibitem{schneider} 
S. Schneider, D. F. V. James and G. J. Milburn, J. Mod. Opt. {\bf 47}, 499 (2000).
%
\bibitem{domokos} P. Domokos, J. M.~Raimond, M. Brune and S. Haroche, 
Phys. Rev. A {\bf 52}, 3554 (1995). 
%
\bibitem{zoller} T. Pellizzari, S. A. Gardiner, J. I. Cirac and P. Zoller, Phys. Rev. Lett. {\bf 75}, 3788 (1995).
%
\bibitem{rausch}
A. Rauschenbeutel, G. Nogues, S. Osnaghi, P. Bertet, M. Brune, J. M. Raimond and S. Haroche, Phys. Rev. Lett. {\bf 83}, 5166 (2000).
%
\bibitem{brennen}
G. K. Brennen, C. M. Caves, F. S. Jessen and I. H. Deutsch, 
Phys. Rev. Lett. {\bf 82}, 1060 (1999).
%
\bibitem{london} A. Beige, S. F. Huelga, P. L. Knight, M. B. Plenio and R. C. Thompson, J. Mod. Opt. {\bf 47}, 401 (2000).
%
\bibitem{briegel}
D. Jaksch, H. J. Briegel, J. I. Cirac, C. W. Gardiner and P. Zoller,
Phys. Rev. Lett. {\bf 82}, 1975 (1999); 
D. Jaksch, J. I. Cirac, P. Zoller, S. L. Rolston, R. Cote, M. D. Lukin,
quant-ph/0004038. 
%
\bibitem{jon}
D. Jonathan, M. B. Plenio and P. L. Knight, Phys. Rev. A (in press);
quant-ph/0002092.
%
\bibitem{steane}
A. Steane, C. F. Roos, D. Stevens, A. Mundt, D. Leibfried, F. Schmidt-Kaler and R. Blatt, quant-ph/0003087.
%
\bibitem{Chuang:95}
I. L. Chuang, R. Laflamme, P. W. Shor and W. H. Zurek, Science {\bf 270}, 1633 (1995).
%
\bibitem{PleKni13} 
M. B. Plenio and P. L. Knight, Phys. Rev. A {\bf 53}, 2986 (1996);
M. B. Plenio and P. L. Knight, Proc. Roy. Soc. London Ser. A {\bf 453}, 2017 (1997).
%
\bibitem{Steane96}
A. M. Steane, Proc. R. Soc. A {\bf 452}, 2251 (1996).
%
\bibitem{Steane962}
P. W. Shor, Phys. Rev. A {\bf 52}, R2493 (1995);
A. R. Calderbank and P. W. Shor, Phys. Rev. A {\bf 54}, 1098 (1996).
%
\bibitem{Zurek83} 
W. H. Zurek, Prog. Theo. Phys. {\bf 89}, 281 (1993).
%
\bibitem{Palma:96}
G. M. Palma, K. A. Suominen and A. K. Ekert, Proc. Roy. Soc. London Ser. A {\bf 452}, 567 (1996).
%
\bibitem{Lidar:PRL98}
D. A. Lidar, I. L. Chuang and K. B. Whaley, Phys. Rev. Lett. {\bf 81}, 2594 (1998).
%
\bibitem{Duan}
L. M. Duan and G. C. Guo, Phys. Rev. Lett. {\bf 79}, 1953 (1997).
%
\bibitem{Zanardi}
P. Zanardi and M. Rasetti, Phys. Rev. Lett. {\bf 79}, 3306 (1997). 
%
\bibitem{Lidar}
D. A. Lidar, D. Bacon and K. B. Whaley, Phys. Rev. Lett. {\bf 82}, 4556 (1999).
%
\bibitem{Duan33}
L. M. Duan and G. C. Guo, Phys.~Rev.~A {\bf 58}, 3491 (1998).
%
\bibitem{exps}
P. Zanardi and F. Rossi, Phys. Rev. B {\bf 59}, 8170 (1999).
%
\bibitem{man}
One method of how to manipulate the states inside a certain DFS, 
which is very different from the approach we propose here, is discussed by D. Bacon, J. Kempe, D. A. Lidar and K. B. Whaley, quant-ph/9908064 and by J.~Kempe, D.~Bacon, D. A. Lidar and K. B. Whaley, quant-ph/0004064. The state of the system always remains completely inside the DFS. This requires an exchange interaction which is not easily available in quantum optics. 
%
\bibitem{others} See also Ref.\cite{Palma:96}; L. Viola and S. Lloyd, Phys. Rev. A {\bf 58}, 2733 (1998); M. Dugi\'c, quant-ph/0001009.
%
\bibitem{knight}
P. M. Radmore and P. L. Knight, J. Phys. B {\bf 15}, 561 (1982).
%
\bibitem{yeoman} 
G. M. Meyer, G. Yeoman, Phys. Rev. Lett {\bf 79}, 2650 (1997).
%
\bibitem{alt} 
M. B. Plenio, S. F. Huelga, A. Beige and P. L. Knight, Phys. Rev. A {\bf 59}, 2468 (1999).
%
\bibitem{meystre} 
G. J. Yang, O. Zobay and P. Meystre, Phys. Rev. A {\bf 59}, 4012 (1999). 
%
\bibitem{misra} B. Misra and E. C. G. Sudarshan, J. Math. Phys. {\bf 18}, 
756 (1977).
%
\bibitem{itano} 
For an experimental test of the quantum Zeno effect see  W. M. Itano, D. J. Heinzen, J. J. Bollinger and D. J. Wineland, Phys.~Rev.~A {\bf 41}, 2295 (1990).
%
\bibitem{behe} 
A.~Beige and G. C.~Hegerfeldt, Phys.~Rev.~A {\bf 53}, 53 (1996);
A.~Beige and G. C.~Hegerfeldt, J.~Phys.~A {\bf 30}, 1323 (1997).
%
\bibitem{HeWi} G. C.~Hegerfeldt and T. S.~Wilser, in
{\it Classical and Quantum Systems.} 
Proceedings of the II. International Wigner Symposium, July
1991, edited by H. D.~Doebner, W.~Scherer and F.~Schroeck: World
Scientific (Singapore 1992), p.~104.
%
\bibitem{HeSo} G. C.~Hegerfeldt and D. G.~Sondermann, Quantum Semiclass.~Opt.~{\bf 8}, 121 (1996).
%
\bibitem{PleKni} M. B.~Plenio and P. L.~Knight, Rev.~Mod.~Phys.~{\bf 70},
101 (1998).
%
\bibitem{MC} J.~Dalibard, Y.~Castin and K.~M{\o}lmer,
Phys.~Rev.~Lett.~{\bf 68}, 580 (1992).
%
\bibitem{QT} H.~Carmichael, {\em An Open Systems Approach to
Quantum Optics}, Lecture Notes in Physics m 18, Springer (Berlin 1993).
%
\bibitem{Mandel} L.~Mandel and E.~Wolf, {\em Optical Coherence and Quantum Optics} (Cambridge University Press, Cambridge 1995).
%
\bibitem{foot}
If the decay rate of the cavity mode is much larger than $g$ the states with
photons in the cavity are emptied quickly and the transition between the
states with $n=0$ and $n=1$ are inhibited by the same mechanism
explained in the previous paragraph. 
%
\bibitem{bose}
S. Bose, P. L. Knight, M. B. Plenio and V. Vedral, Phys. Rev. Lett. {\bf 83}, 5158 (1999).
%
\bibitem{gant} For the general case see e.g. F. R. Gantmacher, {\em Matrizentheorie}, Springer, Berlin (1986).
%
\bibitem{neu} J. von Neumann, {\em mathematische Grundlagen der Quantenmechanik}, Springer, Berlin (1932); G. L\"uders, Ann.~Phys.~{\bf 8}, 323 (1951).
%
\bibitem{CNOT} A. Beige, D. Braun, B. Tregenna and P. L. Knight, submitted to Phys. Rev. Lett., quant-ph/0004043.
\end{references}
\end{document}